\documentclass[11pt]{article}

\usepackage[utf8]{inputenc}

\usepackage[left=2cm,top=1cm,right=2cm,nohead]{geometry}
\usepackage{graphicx} 

\usepackage{booktabs} 
\usepackage{array} 
\usepackage{paralist} 
\usepackage{verbatim} 
\usepackage{subfig} 

\usepackage{fancyhdr} 
\usepackage{sectsty}
\allsectionsfont{\sffamily\mdseries\upshape} 

\usepackage[nottoc,notlof,notlot]{tocbibind} 
\usepackage[titles,subfigure]{tocloft}

\title{\textbf{Activity in the lunar surface: Transient Lunar Phenomena}}
\author{Cruz Roa, Andrés Felipe \footnote{\textbf{}University of Tolima,  Astronomy Group Urania Scorpius, Ibagué - Tolima - 
Colombia, email: afcruz@ut.edu.co}
}
\date{October 1, 2012} 

\begin{document}
\maketitle
\centerline{\textbf{ABSTRACT}}

Transient Lunar Phenomena (TLP) observed on the surface of the moon, are of high rarity, low repetition rate and very short 
observation times, resulting in that there is little information about this topic. This necessitates the importance of studying them 
in detail. 
They have been observed as very bright clouds of gases of past geological lunar activity. According its duration, there 
have been registered in different colors (yellow, orange, red). Its size can vary from a few to hundreds of kilometers. The TLP Usually occur in certain 
locations as in some craters (Aristarchus, Plato, Kepler, etc.) and at the edges of lunar maria (Sea of Fecundity, Alps hills area, etc.). The exposure time of a TLP can vary from a few seconds to a little more than one hour.

\medskip

In this paper, a literature review of the TLP is made to build a theory from the existing reports and scientific hypotheses, trying 
to unify and synthesize data and concepts that are scattered by different lunar research lines. The TLP need to be explained 
from celestial mechanics and planetary astrophysics to explain the possible causes from phenomena such as outgassing moon, moonquakes
and the gravitational interaction. Extrapolating these hypothetical physical knowledge, arguments are exposed for Lunar degassing
theory showing this as the most consistent. It's build also a the theory of how to observe, describe, explain and predict the TLP.

\medskip


\section{INTRODUCTION}
\subsection{What is TLP?}

A TLP is an explosive emanation of  waste gases deposited under lunar soil produced by geological lunar activity, possibly originated 
by a moonquake because of the gravitational pull of the Earth-Moon system. These gases are also influenced by the incident solar  
radiation wich makes them visible.

\medskip

After the formation of the Moon, cooled volcanic parent material formed the lunar maria and gas deposits. Some 3 billion years ago
the moon stopped its geological activity (Stevenson, 1987), but currently NASA researchers have discovered new traces that indicate 
that there have been very recent geological faults (JPL, 2012). Then the heavy bombardment of asteroids and comets took off lunar 
soil stability (Sigismondi, 2000), leaving a fragile scenario for the occurrence of the TLP.

\newpage

\medskip
\subsection{What other examples are in the solar system similar to the TLP?}
\medskip

The same phenomenon occurs in Europe, one of the satellites of Jupiter which is caused by deformation in the crust of the satellite,
as has been observed by the Cassini spacecraft (Hedman, 2009). The gravitational pull of Jupiter and other Galilean moons induce tidal
friction phenomenon whose manifestation is, among others, the formation of small water flows on the surface (Müller, 2007). 
Another example is the satellite Triton, of planet Neptune, releasing geysers of liquid nitrogen which makes it one of the satellites
that have active geological activity. Another case is Enceladus, a Saturn's satellite, with geysers which release particles  in South 
Pole observed as scratches called "Tiger Stripes". These particles make up the E ring of the planet (Matthew, 2010). NASA researchers
determined that "the storms that occur at the poles of Saturn are caused by this phenomenon on Enceladus" (OSH, 2011). 
Therefore, Enceladus would be the only natural satellite of our solar system that influences the chemical 
composition of the planet.

\medskip
\subsection{Ancient and modern observers}
\medskip

Some of the TLP are especially bright because there are reports from the Middle Ages, a time that could only be viewed with a naked eye.
William Herschel (German astronomer who discovered the planet Uranus and hundreds of bodies and observer of many celestial phenomena),
observed these phenomena documenting  and publishing in his "catalogs of double stars and nebulae" (Nasim, 2011), however for 
him was inexplicable.

\medskip

Much later in human history, the Prospector space mission of NASA launched on January 7, 1998 (Cosmopedia, 2008), conducted 
a mapping search for  evidence of ice and recognition of moon's surface features (LPI, 2004). Among his instruments it had a spectroscope 
which detected, in the craters Aristarchus and Kepler, radon emanation from the lunar surface (Taylor, 2003). This was the first 
time the theory of lunar outgassing was proposed (Crotts, 2007), without ruling out the possibility that the moon still present a small
geological activity (Wilhelms, 1987). 

\medskip

Crotts and Hummels, researchers of the Department of Astronomy, Columbia University (CAL, 2011), implemented the robotic monitoring project of
lunar 
images (see Figure 1). For this study the observatories were located on the campus of Columbia University and another at the Cerro 
Tololo Inter-American Observatory in Chile. In turn they have observed and  made better projects such as the study of the TLP, the possible
presence of water in different states of matter and the corroboration that the moon is geologically active, establishing 
statistics and collected data on these events in the lunar surface.

\medskip
\subsection{Theories of TLP}
\medskip

Initially it was thought that TLP had its origin in the active dynamics of atmosphere wich reflects distortions on the Moon and bright
regions.  This theory is consistent with some reports but was questioned by the Apollo XI mission where the crew asked confirmation
reported of TLP by the German Bochum observatory. Neil Armstrong (U.S. astronaut of NASA, was the first man to walk on the moon, and 
observer of a TLP), described the phenomenon near the crater Aristarchus as "an area considerably lighter than the surrounding area"
(Apollo XI, 2010) with a "fluorescence" (property of substances and materials absorb and emit energy in the form of light). This more
comprehensive reporting is important in respect to the description for the study of TLP.

\medskip

A second theory proposes that the TLP are caused by meteorite impacts because the Moon has no atmosphere wich protects against these. 
Never has been found
that a TLP produces a crater. The third theory proposes that the cause is volcanic material spills from the geologic 
lunar activity influenced by sunlight that makes it shine. This theory is not supported by the observations and there is no information to 
justify its role in creating TLP (Anunziato, 2010).

\medskip

The latest theory is that the TLP are caused by lunar outgassing of the gas deposits left from the last geological activity going 
abroad and shine (Crotts, 2008). This theory is more favorability since Lunar Prospector mission of NASA detected gases coming out 
near the craters Aristarchus and Kepler, which served as evidence for the most skeptical and to really be further studied as a 
natural phenomenon happens on the moon (Anunziato, 2010).

\medskip
\subsection{Author's approach}
\medskip

The importance of this research lies in the consolidation of a theory for the study of the TLP, which in turn helps to reinforce the 
theory of lunar outgassing of Arlin Crotts. Having a better theoretical framework, the TLP can be studied in detail. During the 
investigation of the TLP  an exercise is performed try watching one of these rare events (there is no record of the author) and analyze
other images that scientists have managed to capture through observation, demonstrating the practical means of investigating TLP. 

\medskip
\section{WORK SAMPLE}
 \medskip

In the development of this article a concrete theory is established that includes all these data to establish the observation, description,
explanation and prediction of TLP based on the theories and observations of different authors, through past and present records about
the TLP (Bartneck, 2007).

\medskip

\subsection{THEORY}
\medskip
\subsubsection{Observation and description of TLP}
\medskip

This lunar phenomenon  is observed in a cloud shape very highlighted and changing its color visibility in time, since the gases 
interact with the incident sunshine which makes it turn red, yellow, orange, etc. Its shape and dynamics are irregular and inaccurate, 
fading in a few seconds and suddenly (Holmstrom, 2011). Crotts and Hummels have monitored the Moon robotically through images. 
An example of TLP is observed through astrophotography in Figure 1.

\begin{figure}[h]
\centering
\includegraphics[width=380pt]{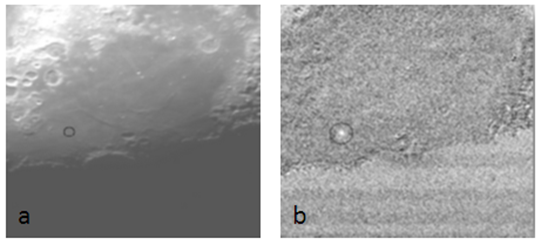}
\end{figure}

\captionof{figure}{Transient Lunar Phenomena in the circle: a) raw image and b) processed image. (Crotts and Hummels, 2010).}

\subsubsection{Explanation of TLP, from the theory proposed}

The explanation of TLP is not fully developed. As mentioned, one of the most consistent theories is lunar outgassing (Crotts, 2009), 
because the Apollo missions revealed the possibility that the moon had a geological activity until a million years ago showing 
evidence such as INA structure, this is a zone of craters and mountains are in the latest evidence of this satellite orogenic 
activity (see Figure 2). These gases were stored in deposits, due the Moon  presents a cortex structurally homogeneous 
(Phillips, 2006).

\begin{figure}[h]
\centering
\includegraphics[width=300pt]{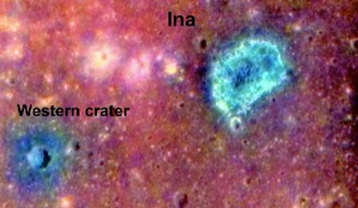}
\end{figure}

\captionof{figure}{INA Structure, evidence of recent geological activity (one million years). (Phillips, 2006).}

\medskip
\medskip
\medskip
\medskip

\textbf{Illustration}
\bigskip

Theoretically, the Moon formed as a result of an impact between primitive Earth and a body about one seventh the size of Earth 
(Ziethe, 2008), also known as Theia (the moon goddess in Greek mythology), which disintegrated and formed the Moon. As a result 
of the collision, temperatures up to 10.000 ºC were reached,  which resulted in melting and evaporation of materials present in the 
two bodies, producing that large fragments of primitive crust were ejected into space near the Earth; then the natural process of 
accretion (addition of material to a body in astronomical time is a celestial body larger) generated naturally our satellite 
(Montgomery, 2009).

\medskip

Then, this orbits system achieved its stability determining the rotation of the earth, as the earth originally had a turnover of 6 hours
(Fahrenholz, 2011). The influence of the moon slows the rotation down to be approximately 24 hours (Iorio, 2011). 
Thanks to the planet Earth has a natural satellite of large diameter (3,473 km), mass ($ 7.35 x 10 ^ {22} $ kg) and density 
(3.34 g/cm3) (Hack, 2011) these were equilibrated mutually and this is how we see them today.

\medskip

During the geologic lunar activity in this period of time, TLP were produced suddenly and in large quantities (Crotts, 2009). 
Internal and external flows of magma are ejected and the process is repeated cyclically. In the course of solidifying. the magma 
flows  created cavities filled with gas and steam (Bratkov, 2006). Superficially, the relief was defined temporarily  (Gaddis, 2002). 
These spots called the attention of one of the most important and brilliant astronomers in history, Galileo Galilei that he named
"lunar maria", based on the shape, they look like the seas of the earth viewed from space. Thereafter, the study of dark spots present 
on the lunar surface is known as Selenography which is the study of surface characteristics and physical properties of the moon 
(Russell, 2011).

\medskip

These places on the moon remained for many years in stable geological and topographical, but natural events as rain of meteors, 
meteorites, comets and asteroids from the asteroid belt, the Kuiper belt and the Oort cloud (region of space occupied by rock 
fragments frozen material and having a means of dust and gas located in the outer limits of our solar system) (Pieters, 2008), 
hit the surface without control, as the despicable lunar atmosphere has a low coefficient of friction that fast enough to disintegrate
the bodies (see Figure 3), forming the craters on the moon that we can detail (Sigismondi, 2001).

\medskip
\medskip
\medskip

\begin{figure}[h]
\centering
\includegraphics[width=300pt]{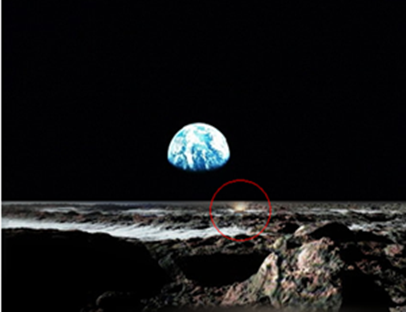}
\end{figure}

\captionof{figure}{Viewing a pre-stage for the formation of a transient lunar phenomenon. It represents the lunar surface in the 
in interaction with the Earth. The red circle indicates cavities-storing waste gases from the past geological activity.
(modified of http://www.bitacoradegalileo.com/, 2011).}

\medskip
\medskip
\medskip
\medskip
\medskip
\medskip

The Moon is under the influence of celestial bodies as the Earth, the Sun, planets and others. As a first step, the Moon is subject 
to the gravitational pull of the Earth, which together with the eccentricity of the orbit, causes that such interaction is variable, and 
 its intensity is extreme both at apogee (farthest point from the Earth the moon, 406,740 km) and perigee (nearest point between 
the Earth and the Moon, 356.410km) (Portilla, 2011), which may cause  moonquakes  causing slight surface deformation. The deformation 
disturbs materials, gas deposits and water vapor ejecting from the depths of the lunar soil outside (Nayakshin, 2008). This lunar 
phenomenon 
 occurs very frequently in certain craters due to the nature of their training, as they are more unstable and eroded that lunar 
maria; harder and solidified. Because these interactions,  fractures or small explosions are created which in turn serve as tubular channels
that funnel gases, which are sporadically released from deep in the crust of the craters 
or directly from the lunar maria (see Figure 4) (Crotts, 2010).

\medskip

These gases are affected by the sunshine, that sometimes on the surface of the moon can be increased to  200 ºC at high solar 
activity (Holmstrom, 2010). This high temperature causes the gases become fluorescent as they are at a temperature below 0 ºC, so 
after leaving such deposits and receive the bombardment of sunshine incident, they are heated and change their color and interact with other 
gases forming a cloud of different gases (Crotts, 2008). On the other hand, for the lunar maria can produce TLP, a moonquake oh high proportion 
is required that may cause fractures which residual gas escapes Lunar geological activity (Berezhnoi, 2000).

\newpage

\begin{figure}[h]
\centering
\includegraphics[width=300pt]{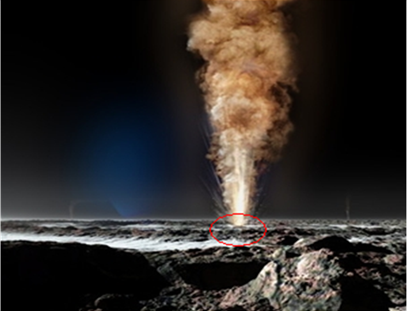}
\end{figure}

\captionof{figure}{Viewing a transient lunar phenomena. The red circle indicates a deposit of waste gases, that are emanating and 
interacting with the incident sunshine. (Modified http://www.bitacoradegalileo.com/, 2011).}
\bigskip
\bigskip
\bigskip

\textbf{Active craters in transient lunar phenomena}
\bigskip

The craters that have submitted reports of TLP are the craters Aristarchus presenting  about 42 strong reports, with 13 entries 
Plato, and Kepler with 3 records. They are the most active and reported  craters by the observers (Crotts, 2008). Some results established 
that "about 50 \% of the reports originally occurred in the crater Aristarchus, 16 \% in the crater Plato and 6 \% in recent craters
(Copernicus, Kepler and Tycho)" (Crotts, 2007) . Table 1 shows the strong reports analyzed at different intervals of time, which 
includes other craters having at least one blunt report.

\bigskip 
\bigskip 

\captionof{table}{Number of reports of transient lunar phenomena for three historical periods}

\bigskip 

\begin{figure}[h]
\centering
\includegraphics[width=430pt]{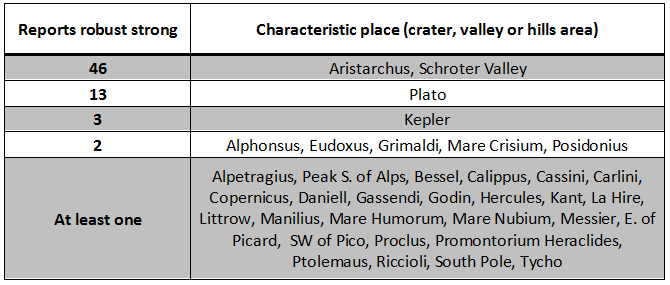}
\end{figure}

Source: Crotts (2009).

\newpage

\textbf{Prediction of TLP}
\bigskip

In principle, the FLT can be predicted by knowing the time when the Earth and the Moon are closer between them, where the
gravitational attraction is greater. For this reason, it can cause multiple moonquakes in the months of January, 
February, August and September as shown in the Figure 5. Because the moon is under constant influence of the sun, there are 
disturbances in the orbit, causing the distances (perigee and apogee) vary with time.

\medskip

Furthermore, the effect is more pronounced in the presence of high solar activity as flares or coronal mass ejections (Datta, 2008),
including the influence of fireballs (frozen fragments, meteorites, comets and in some cases asteroids)  that impact the moon and 
produce moonquakes and elevated temperatures (Iorio, 2011). Observing the top 8 craters of the Table 2 could probably spot a TLP 
with the above variables using Figure 6 as a tool of the researcher Arlin Crotts.

\bigskip
\bigskip
\bigskip
\bigskip

\begin{figure}[h]
\centering
\includegraphics[width=380pt]{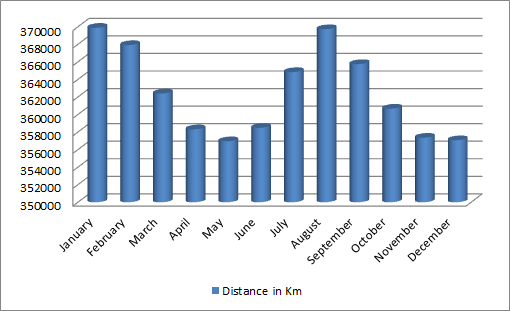}
\end{figure}

\captionof{figure}{Apogee lunar month of the year, gravitational pulls more intense. (Source: The author)}

\newpage

\captionof{table}{Relative frequencies of reported transients lunar phenomena in percentage}

\begin{figure}[h]
\centering
\includegraphics[width=360pt]{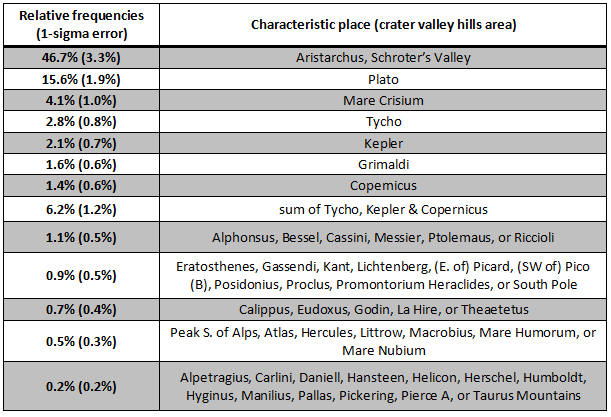}
\end{figure}

Source: Crotts (2009).

\begin{figure}[h]
\centering
\includegraphics[width=290pt]{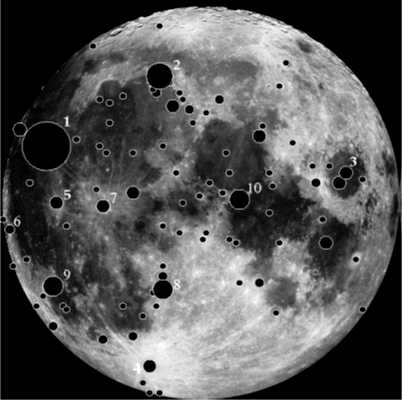}
\end{figure}

\captionof{figure}{Distribution of transient lunar phenomena based on reports, with the exception of a minority of cases are 
rejected by the conditions of truthfulness. The size of the circles represents the number of reports of transient lunar phenomena: 1)
Aristarchus (even Schröter Valley, the Cobra Head and Herotus), 2) Platón, 3) Mare Crisium, 4) Tycho, 5) Kepler, 6) Grimaldi, 7) 
Copernicus, 8) Alphonsus, 9) Gassendi and 10) Ross D. (Taken NASA, quoted by Crotts, 2009).}

\bigskip
\bigskip

\subsection{PRACTICAL ASPECT}
\medskip

In the present research practice, lunar images were captured in monthly time intervals and usually  in waxing and waning moon phases,
because it gives a better detail of the lunar surface unlike full and new moon phases .The observations were made using 
a 60 mm refractor telescope, a Genius digital camera of 6 mega pixels, a PC and the software Astroart that 
allows to process and make 3D visualizations of the active craters in the TLP. What is sought by means of the captured scene is a small
salient point to the other details of the Moon (see Figure 7a) and then be processed and analyzed by the software. The 3D images
displayed a protuberance lighter than the surrounding area (see Figure 7b), in this way, by capturing lunar images  we can analyze and 
determine whether TLP occurred at that instant of observation.

\bigskip

The observationt are shown in the images captured on Sept. 11, 2011, on the outskirts of the
Ibague city at 01:23 UTC (8:23 pm). The monitoring was mainly focused on the active craters such as Aristarchus, Plato and Kepler, as 
these are craters that have large number of records and probability of sighting a TLP.

\medskip

\begin{figure}[h]
\centering
\includegraphics[width=380pt]{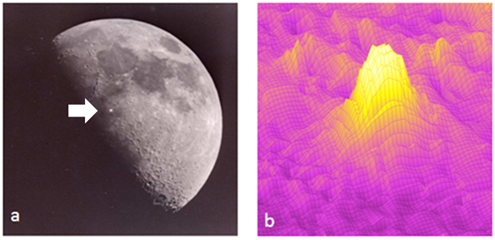}
\end{figure}

\captionof{figure}{a. Photography a TLP. It shows a small point in the center of the image. (source: Stuart, 1953). b. Image
processed in Astroart, showing the light having the small point of the original image. (Source: The author).}

\bigskip
\bigskip

Schröter Valley, the plateau and the crater Aristarchus including Herodotus (see Figure 8a) has a shape similar to a mound of 
mountain building result that simultaneously serves as a channel for tubular exhaust (Wieczorek, 2006), as seen on Earth, 
presenting geological activity and lava is escaping from the inside . One example are the geysers of Yellowstone park in the United States. 
This valley, plateau and crater are the main subject of comparative study of the TLP.

\medskip

The crater Plato (see Figure 8b) has a flat surface and has a high number of registers TLP. In turn, it covers a vast area of soft land  
that when is disturbed with small moonquakes can form fractures that serve as escape routes of gases and water vapor,  because this region 
can accommodate ice thanks to the low temperature in the dark of the lunar night (Boersma, 2009). There are very old reports of the 
presence of TLP in this crater.

\medskip

The crater Kepler (see Figure 8c) is also active in the TLP, therefore is not surprising that the Prospector spacecraft detected 
gases emanating from there (Kobrick, 2010). In this crater some minerals produce gases which then rise to the surface; they 
shine because the incident light of the sun and form a possible TLP (Shuster, 2009).

\newpage

\begin{figure}[h]
\centering
\includegraphics[width=450pt]{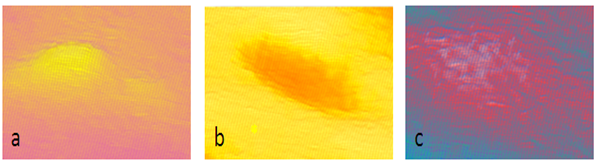}
\end{figure}

\captionof{figure}{Active craters in the TLP: a) Plateau and Aristarchus crater, palette color, b) Plato crater, flame palette, 
c) Kepler crater, jazz palette. Images captured by a telescope 6cm of diameter with 70 cm focal length, Genius digital camera 
and processed Astroart software. (Images captured and processed by the author).}

\section{DISCUSSION}

The main current uncertainty study of TLP are the instruments by which to capture images that are likely to have an optical defect,
such as noise when capturing the scene; but it is remarkable that there are reports and hypotheses erroneous against reports forceful
as shown in Figure 9 wich makes clear that the phenomenon actually occurs on the surface of the moon, not because of the photographic 
instrument. Another plus on the accuracy of these phenomena is that if  noise would be displayed sparsely, but these protuberances 
or risers are well defined and do not spread throughout the picture.

\begin{figure}[h]
\centering
\includegraphics[width=350pt]{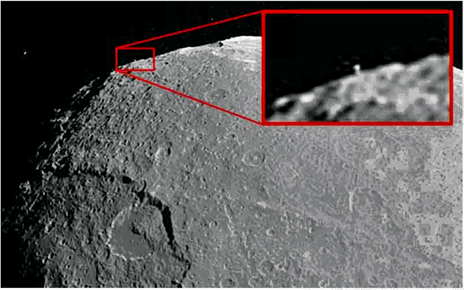}
\end{figure}

\captionof{figure}{Image a TLP. In the scene shown a small riser, front view. (source: Darnaude, 2010).}

\bigskip

The observations and theoretical work suggests that it is possible that the Sun-Earth-Moon system exert disturbances influence
in the TLP and lunar outgassing theory is the most likely explanation, since this includes and explains the slight manifestations
of geological lunar activity causing the TLP. The Astronomer Bonnie Buratti, a researcher at the Jet Propulsion Laboratory and NASA's,
is one of the people who have studied this and other lunar phenomena (JPL, 2011). She states that "no convincing
observers to visualize the TLP and so data can be rectified and convincing explanation about transient lunar phenomena" (Long, 2007).
Therefore, further study should be made of our natural satellite, to verify the theories that have been formulated (Meijer, 2008). 
It is hoped that this work will contribute to this verification of the theories and observations of the TLP.

\bigskip

\section{CONCLUSIONS}

According to the theory proposed in this article for the TLP, the moon possibly should have a reduced  geological activity on the
basis of the recent INA structure and TLP. Of course, these events are in smaller proportions to those in the earth and planetary 
systems studied: the case of  Europea of Jupiter, Saturn's Enceladus and Neptune's Triton. It is explained that in theory are mainly
radon gas deposits influenced by geological events and interacting with sunlight. 

\medskip

It is Analyzed the existence of a greater chance to spot, record, photograph and study a TLP. Using consolidated statistics Arlin Crotts 
(2007) states that 80 \% of the sightings are real. Cameron Hummels (1972), stated that there are about 771 reports of TLP in 
different catalogs, giving the main targets for shooting with the telescope and then process these images using the Astroart software.

\medskip

In this work is mainly accepted the characteristics and parameters of the lunar outgassing theory proposed in the articles of Arlin Crotts, 
as it is probably the right one for the explanation of the TLP. It should be noted that the phenomena are related to craters 
with the lunar maria, the most active are the areas relating to the crater Aristarchus, and younger craters.

\medskip

Space missions such as the launched in September 10, 2011 GRAIL (Gravity Recovery and Interior Laboratory) will provide more 
information of the selenography of inside and outside of the moon and so help to test different theories about of the TLP to gain 
more scientific knowledge of the natural satellite of the planet Earth.
\bigskip
\bigskip 
\bigskip 
\bigskip 

\textbf{ACKNOWLEDGEMENTS}
\bigskip

The author is grateful to PhD José Herman Muñoz of Department of Physics, University of Tolima; teachers Benjamin Calvo and 
José Gregorio Portilla of Astronomical Observatory of the National University; Professor Alberto Quijano Vodniza the Astronomical
Observatory of the University of Nariño; Professor Alejandro Guarnizo of Heidelberg University, Germany; to Mr. Julio Ernesto 
Paipa Nieto;  Astronomy Group Urania Scorpius in memory of Professor q. e. p. d. Alonso Medina, for his guidance and academic
counseling.

\bigskip \bigskip \bigskip\bigskip \bigskip\bigskip\bigskip \bigskip \bigskip\bigskip \bigskip \bigskip\bigskip\bigskip
\bigskip \bigskip \bigskip\bigskip \bigskip\bigskip\bigskip \bigskip \bigskip\bigskip \bigskip \bigskip\bigskip\bigskip

This preprint was prepared with the AAS  \LaTeX\  macros v5.0.

\end{document}